\documentclass[letterpaper]{sfchem}
\usepackage{graphicx}

\begin{document}

\title{The Serpens Star-Forming Region in HCO$^+$, HCN, and N$_2$H$^+$}

\author{Michiel R.\ Hogerheijde} 
  \institute{Steward Observatory, The University of Arizona, 933 N. Cherry
  Ave, Tucson, AZ 85721, U.S.A.} 
\authorrunning{Hogerheijde}
\titlerunning{Serpens in HCO$^+$, HCN, and N$_2$H$^+$}

\maketitle 

\begin{abstract}
This poster presents single-dish and aperture-synthesis observations
of the $J$=1--0 ($\lambda\approx 3$~mm) transitions of HCO$^+$, HCN,
and N$_2$H$^+$ towards the Serpens star-forming region. Jets driven by
young stars affect the structure and the chemistry of their
surrounding cloud, and this work aims to assess the extent to which
the emission of these three molecular lines is dominated by such
processes. In Serpens I find that N$_2$H$^+$ 1--0 traces the total
amount of material, except in two regions slightly ahead of shocks. In
contrast, the HCO$^+$ and, especially, HCN emission is dominated by
regions impacted by outflows. One previously unknown, strongly shocked
region is located $\sim 0.1$ pc northwest of the young stellar object
SMM~4. There is a marked spatial offset between the peaks in the HCN
and the N$_2$H$^+$ emission associated with shocked regions. I
construct a simple, qualitative chemical model where the N$_2$H$^+$
emission increases in the magnetic precursor of a C-type shock, while
N$_2$H$^+$ is destroyed deeper in the shock as the neutrals heat up
and species like HCN and water are released from icy grain mantles.  I
conclude that N$_2$H$^+$ is a reliable tracer of cloud material, and
that unresolved observations of HCO$^+$ and HCN will be dominated by
material impacted by outflows.


\end{abstract}

\section{Introduction\label{s:intro}}

The formation of stars is accompanied by energetic activity such as
jets and outflows. This can affect the structure and the chemical
composition of the surrounding cloud, possibly influencing its
continued star formation. A relevant question therefore is, to what
the extent the emission of commonly used tracers of dense gas reflects
the energetics of star formation rather than the underlying structure
of the cloud. This is especially interesting for unresolved
observations of clouds and cloud complexes, at large distances in our
Galaxy or in other galaxies (e.g., \cite{aalto95,helfer97,wild00,kuno02}). This poster contribution investigates the emission lines of
three molecules (HCO$^+$, HCN, and N$_2$H$^+$) in the Serpens
star-forming region, with the specific question in mind of the
relation between the emission and activity of young stars.

The Serpens star-forming region is ideally suited for the stated
aim. It is relatively nearby at $\sim 310\pm 40$ pc (\cite{delara91})
and harbors several deeply embedded protostars as well as apparently
starless dust condensations (\cite{casali93,davis99}). The molecular
cloud consists of two subcondensations, northeast (NE) and southwest
(SW), each of which are broken up into numerous subclumps and
filaments. Many outflows emanate from the Serpens region
(\cite{davis99}), apparently driven by the embedded protostars. The
high density of young stellar objects (YSOs) and the complex structure
of the region precludes identification of the driving sources of many
of these flows. It is likely that the course of some of the flows are
altered by collisions with dense cloud material, further confusing the
picture.

\begin{figure*}
\centering
\includegraphics[width=0.8\linewidth]{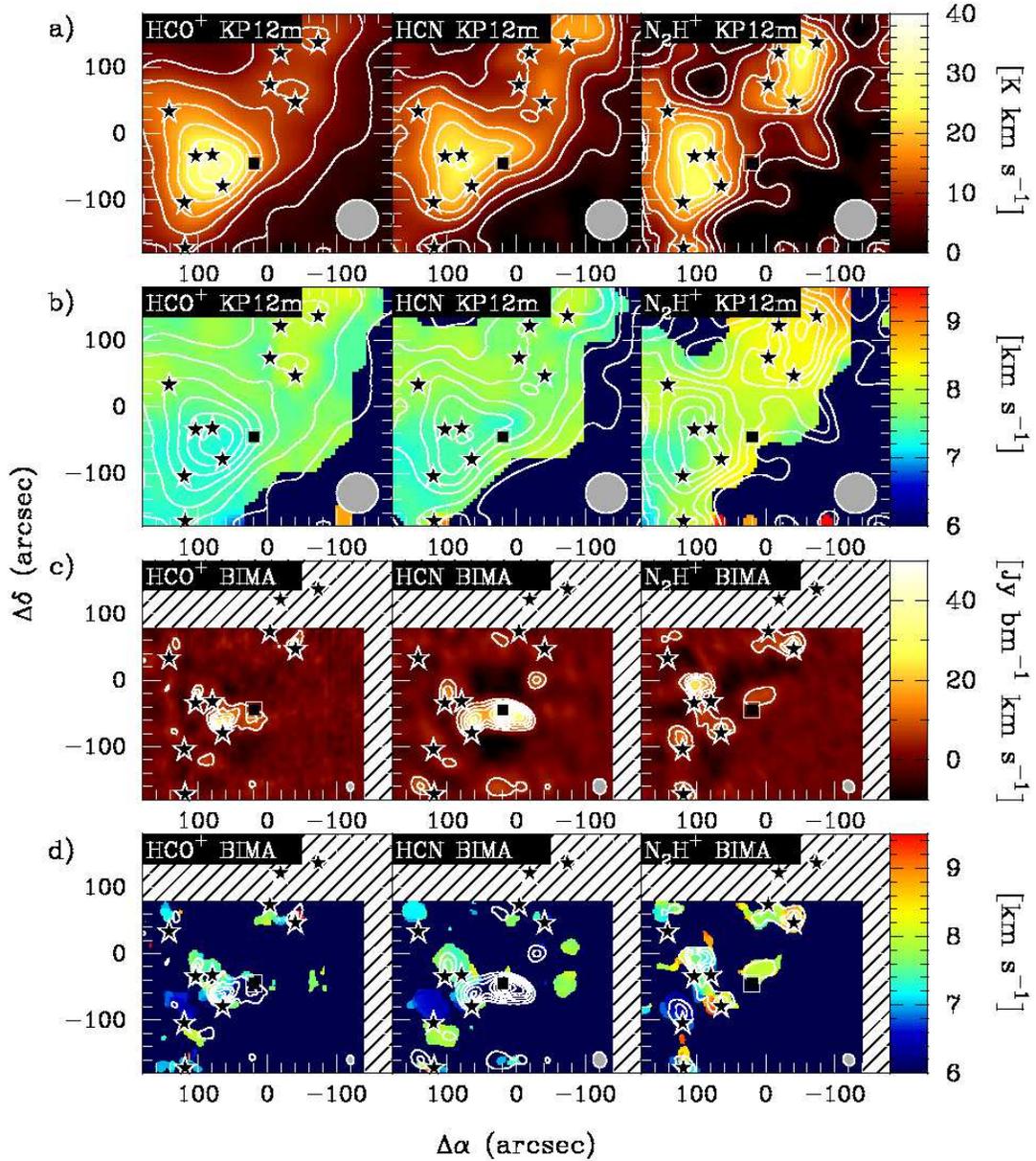}
\caption{(a) KP12m maps of integrated intensity. (b) KP12m maps of the
velocity centroid, with contours of integrated intensity
superposed. (c) BIMA maps of integrated intensity. (d) BIMA maps of
velocity centroid, with contours of integrated intensity
superposed. The star symbols mark the locations of the YSOs; the
square marks the location of the `shock front'. The beam sizes are
represented by the grey ellipses in the lower right-hand corner of each
panel. The BIMA maps only cover the SE region.\label{fig1}}
\end{figure*}

\section{Single-dish observations\label{s:kp12m}}

In this poster I present single-dish observations of the $J$=1--0
lines of HCO$^+$, HCN, and N$_2$H$^+$ obtained with the Kitt Peak
12-meter (KP12m) telescope\footnote{The 12m Telescope is a facility of
the National Science Foundation currently operated by the University
of Arizona Steward Observatory under a loan agreement with the
National Radio Astronomy Observatory.}; aperture-synthesis
observations are discussed in \S\ \ref{s:bima}. The KP12m maps cover a
$420''\times 420''$ area encompassing the NE and SW condensations
(Fig. \ref{fig1}). The beam size of the telescope is $60''$. The
N$_2$H$^+$ emission traces the NE and SW condensations equally,
resembling the SCUBA 850 $\mu$m emission that traces cold dust
(\cite{davis99}). The HCN and HCO$^+$ emission is dominated by the SE
condensation, where most of the YSOs are located and where most of the
outflows originate. This indicates that N$_2$H$^+$ reliably reflects
the distribution of the cloud material, while HCO$^+$ and HCN emission
is enhanced near outflows. The velocity centroids of the emission
lines show an east--west gradient of 2.5 km~s$^{-1}$, most clearly
seen in N$_2$H$^+$, indicating solid-body rotation of the entire
cloud, as reported by \cite*{olmi02}.

\section{Interferometer observations\label{s:bima}}

To investigate if the close match between N$_2$H$^+$ and dust
continuum, and the association of HCN and HCO$^+$ with outflows holds
on smaller scales, aperture-synthesis images are shown in
Fig. \ref{fig1}. These interferometer observations where obtained at
the BIMA\footnote{The BIMA array is operated by the Universities of
California (Berkeley), Illinois, and Maryland, with support from the
National Science Foundation.} millimeter array, at resolutions of
$12''$--$21''$. Mosaics of 13 array pointings cover a $320''\times
260''$ region around the SE condensation; the NW condensation has
already been studied at high angular resolution by \cite*{williams00}.

There exists an almost one-to-one correspondence between the peaks in
the N$_2$H$^+$ BIMA map and the 850 $\mu$m emssion. The only
excpetions are two N$_2$H$^+$ peaks, one north of SMM~3 and one $\sim
60''$ northeast of SMM~4. As discussed below, both locations probably
are shocks. The association of HCO$^+$ and HCN with outflows
become very clear at the small scales traced by BIMA. Emission is
found north of SMM~3 and SMM~4, along the directions of the jets of
these sources. An emission peak, especially prominent in HCN, is
located $\sim 60''$ northwest of SMM~4. The velocity-centroid of the
emission around this peak stands out as extremely blue, at $>2$
km~s$^{-1}$ from systemic. The likely interpretation of this peak is a
`shock front' of a jet impacting a cloud fragment, although no obvious
driving source is present. Possible candidates are a jet from (the
vicinity of) SMM~1 or from (the vicinity of) SMM~4. In the latter case
it may be argued that at the `shock front' this jet deflects off a cloud
condenstation and later connects up with the outflow lobe marked `Wr'
in the CO map of \cite*{davis99}.

Interferometers are only sensitive to small-scale emission, and filter
out emission on spatial frequencies below the smallest antenna
separation. Compared to the KP12m data that do cover these scales, the
BIMA data contain $\sim 30$\% of the flux. While this is only a small
fraction of the emission, the similarity in distribution of the
emission in the KP12m and BIMA maps suggests that outflows contribute
significantly to the total emission of HCO$^+$ and HCN. Hogerheijde
(in prep.) analyses the emission and the abundances quantitatively.

\section{A shock model for HCN and N$_2$H$^+$\label{s:shockmdl}}

Close inspection of the HCN and N$_2$H$^+$ BIMA maps near the tip of
SMM~3's jet and near the `shock front' reveals that N$_2$H$^+$ peaks
$\sim 20''$ ahead of HCN. The structure of C-type shocks
(\cite{draine93}) may offer an explanation for this offset. C-type
shocks have a magnetic precursor where charged particles accelerate
and warm up ahead of neutral particles, enhancing the intensity of
N$_2$H$^+$ emission lines. Deeper into the shock the neutrals
accelerate and warm up. If the pre-shock density is sufficiently high,
dust grains will be predominately neutral, and it is in this region
that they will release the contents of their ice mantles. These
include HCN, as seen toward other outflow (\cite{lahuis00,mrh01}),
water, and CO. The latter two effectively destroy N$_2$H$^+$ through
chemical reactions, causing the N$_2$H$^+$ emission to disappear where
the HCN peaks.  A quantitative model has to include time scales,
length scales, and an adequate chemical network, and is discussed in
Hogerheijde (in prep.). High-angular resolution observations with,
e.g., the Smithsonian Millimeter Array, can help test this model.

\section{Conclusions\label{s:conclusion}}

Aperture-synthesis and single-dish maps of the Serpens star-forming
region show that N$_2$H$^+$ accurately traces the distribution of the
cloud material, while HCO$^+$ and especially HCN probe material that
is shocked by outflows. Unresolved observations of the latter two
species therefore would reflect star-forming activity rather than
cloud column density. A difference map of N$_2$H$^+$ and HCN may be a
useful tool to identify regions in cloud complexes that are actively
star forming, as opposed to dense but quiescent. On small scales,
however, N$_2$H$^+$ emission may be enhanced in magnetic precursors of
C-type shocks, but appears effectively destroyed in the warmest
regions of shocks where HCN peaks.

%

\end{document}